# On model architecture for a children's speech recognition interactive dialog system

**Radoslava Kraleva, Velin Kralev**
*South-West University "Neofit Rilski", Blagoevgrad, Bulgaria*

**Abstract**: This report presents a general model of the architecture of information systems for the children's speech recognition. It presents a model of the speech data stream and how it works. The result of these studies and presented veins architectural model shows that research needs to be focused on acoustic-phonetic modeling in order to improve the quality of children's speech recognition and the sustainability of the systems to noise and changes in transmission environment. Another important aspect is the development of more accurate algorithms for modeling of spontaneous child speech.

## 1. INTRODUCTION

Technologies for speech recognition are oriented towards the speech recognition of adults. Developed applications, the justification of these technologies are aimed primarily at elderly consumers. Recently children have become users of more and more automated devices like computer equipment, household appliances or toys. Under certain conditions the application of certain technologies can be successful, but in most cases they are hampered by major limitations in the treatment of children's speech. This is because there are too many differences in speech between adults and children. Most of the children are inexperienced in the formulation of words and sentences. It often happens, when the child is in its early age, he/she to replace one phoneme or word for another, to miss one up to several phonemes. Many children use the limited vocabulary of adults. They unlike adults have very strong imagination and association skills that help them create their own new words. Therefore, the dictionary used by children is very different from that of adults.

Therefore, it may be argued that the creation of technology to be applied in children's speech recognition is an interesting and easy task. There are many studies on this matter. The authors are interested in testing the systems for the recognition of child speech in English.

This article proposes a model of information system architecture for children's speech recognition.

## 2. MODEL ARCHITECTUE FOR CHILDREN'S SPEECH RECOGNITION SYSTEM

A speech recognizer might be divided into four parts: the digitizing, feature extraction, acoustic modeling, and language modeling part. When the signal is digitized an appropriate sampling frequency needs to be chosen to catch the high-pitched voices of children. Feature extraction often involves approximating the spectrum of the signal. During this phase it is possible to use signal processing to alter the spectrum prior to recognition. This may be used to decrease the sensitivity against background noise or to normalize some spectral characteristics of the recorded speech. Another method to target the current spectral characteristics may be to adapt the acoustic models to better match the speech. Adjustment of the language model may be used to take the children's vocabulary and pronunciation into account [1]. Adaptation may be performed on many levels. Adaptation has become linked with altering the acoustical model in the speech recognizer.

According to [2] and [3] using an adult speech recognizer without any adjustment to the needs and peculiarities of children speech gives a high word error rate. In particular, speech recognition in very young children becomes much more difficult problem to solve rather than speech recognition in older children. Therefore, the adaptation of existing methods and technologies working well at adult speech recognition is an often used practice with children's speech recognition.

Here we will present a model of the architecture of a speech recognition system which is based on classical models and adjusted to children speech recognition.

### 2.1. *Simple analysis for speech recognition*

Before we move on to presenting the architecture model of a children's speech recognition system, let us first examine the work of recognizers. Knowledge from various scientific fields such as mathematics, physics, acoustics, linguistics, etc. is necessary for the quality performance of this recognition process. This multiple knowledge level processing is still not well studied and this fact is the main reason for the automatic speech recognition to become a complex and not yet solved problem. However, there are a number of practical implementations, which in one way or another have managed to cope with this problem. Schematic summary presentation of the process of recognizing spoken language can be seen in Figure 1

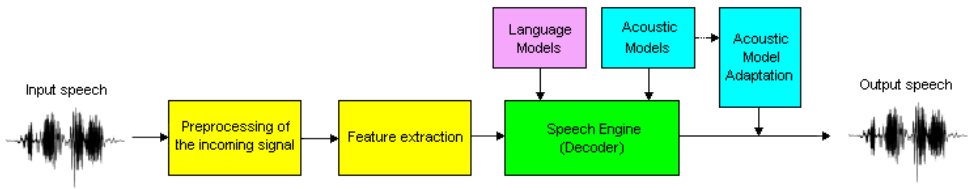

Fig. 1: Schematic structure of the basic steps of an automatic speech recognition system

First, the waveform received is preprocessed to enhance the signal by means of simple preemphasis filtering, or more sophisticated noise reduction or dereverberation algorithms. Then, the processed speech waveform is converted into a sequence of meaningful feature vectors, which are used to feed the speech decoder. The decoder seeks for the best match between a sequence of features and every possible sequence of acoustic classes, using the available information from the acoustic and language models, which are typically obtained in a training phase prior to the recognition step. Additionally, the recognition result can be used as feed-back information to better adjust the acoustic models to the actual speaker or acoustic environment in a process generally named adaptation.

### 2.2. Description of the model architecture

The proposed architecture (figure 2) consists of five main components which are designed to work together. Modules *"Language model"* and *"Dialogue Manager"* can be used and integrated into other independent systems. Module "*Preprocessing of the incoming signal*" includes filters to clear the noise from the received signals or due to imperfections in the sensors. After this the cleared analogue signal is being converted into a digital one by means of an analogue/digital converter.

The "*Speech recognition*" module is performed by the formation of many characteristic features of object recognition by input digital data flow. Then a description of classes and data necessary to perform the classification (decoding) in the "*Classification*" block is included. This block is connected to the *Speech Database*, through which training, self-organization and recognizing words. After this the speech is converted into text.

Understanding natural language recognizes the importance of words and phrases in the context they are used. If the word is not recognized, then it returns to "*Classification*" block and is recorded in the database as associated with keyword depending on its context.

Interpretation and response to those carried out in the next module of the "*Automatic generation of speech*." The tree of conclusions is determined in the "*Speech interpreter*" block. Then, the information which must be re-turned as a response to the user of the system is determined in the "*Response generation*" block. The "*Acoustic model*" block generates the phonetic representation of the speech signal response of the system. In developing the software, it will be selected and modified some existing methods of acoustic modeling, which will best match the characteristics of the Bulgarian language used by children.

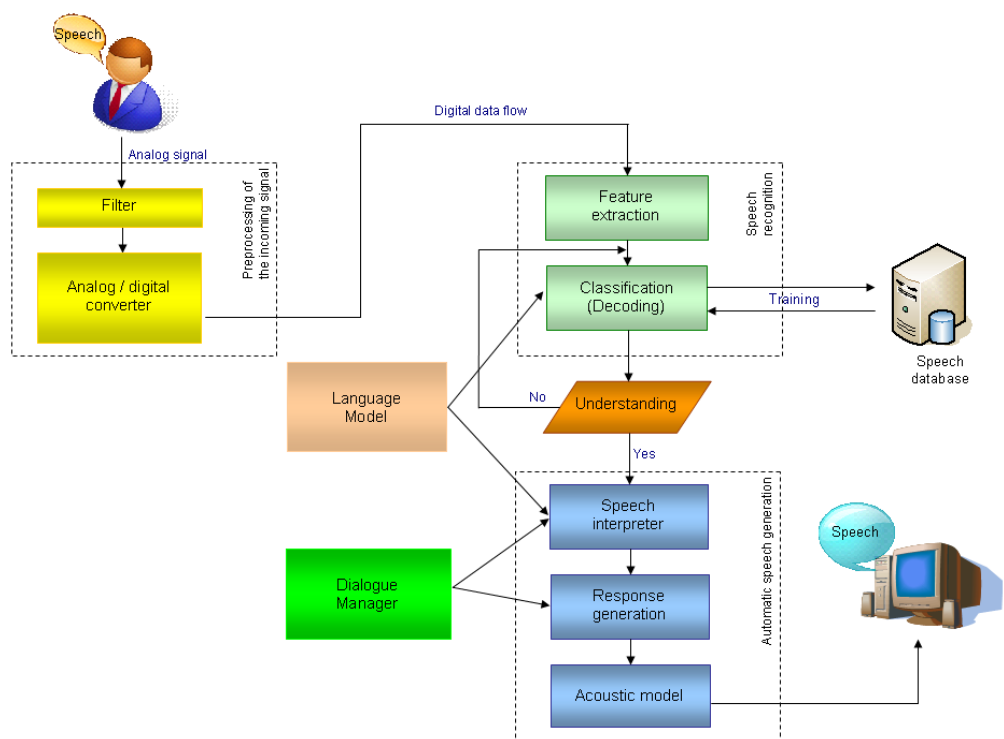

Fig. 2: Schematic structure of the basic steps of an automatic speech recognition system

The "*Dialogue manager*" module controls "*Automatic speech generation*" operations. The most appropriate response based on information submitted to the block appears here. Thus creating a prerequisite for conducting a real dialogue and retain its history. "*Dialog manager*" according to [5] can be seen as a solution to two specific problems: (1) providing a coherent overall structure to interaction that extends beyond the single turn, (2) correctly manage mixed-initiative interaction, allowing users to guide interaction as per their (not necessarily explicitly shared) goals while allowing the system to guide interaction towards successful completion.

Organization of the "*Language model*" will correspond to the peculiarities of the vocabulary of Bulgarian.

Organization of *Speech database* will be similar to the one used by the team of [4]. Separate words and short sentences covering all the features of the phoneme in Bulgarian will be kept. Recordings of the words will be made by a desktop microphone to eliminate the need of expensive and specialized equipment. Data set will be divided into sets for training, development and evaluation. Usability of the words will be determined by the frequency dictionary proposed by Botseva [6]. The record of the words will be made by English children in different age groups from 2 - 7 years of age and without speech defects. Each child will record 80-100 words, as their number will be determined later. The data base formed in this way will be used for training, testing and evaluation.

## 3. CONCLUSIONS

The biggest advantage of the proposed model is its universality. This is because the *Dialog Manager* and *Language Model* are not dependent on the task being executed. They are separate modules that can be updated independently of the system.

The following tasks were set in this report:
- The process of recognizing speech was studied and presented.
- It was analyzed the specifics of the problem associated with the recognition of child speech.
- It was presented an architectural model of a system for recognition of child speech.

Future work:
- A prototype of an information system using the presented model, in order to study the problems related to children's speech recognition in children up to 7 years.
- A language adaptation of existing models will be carried out in order to meet the limited vocabulary used by children.
- • An adaptation of acoustic models and techniques will be carried out to meet the specifics of the speech in children

## 4. REFERENCES

[1] Elenius D., (2004) Adaptation techniques for children's speech recognition, Course in Speech Recognition, Royal Institute of Technology, Sweden

[2] Gustafson, J. and Sjolander, K., (2002) Voice transformations for improving children's speech recognition in a publicly available dialogue system, In the Proceedings of the International Conference on Spoken Language Processing 2002, pp 297 - 300.

[3] Narayanan, S. and Potamianos, A., (2002) Creating conversational interfaces for children. IEEE Transactions on Speech and Audio Processing, Volume: 10, Issue: 2, Feb. 2002, pp 65 – 78
[4] Oppelstrup, L., Blomberg, M. and Elenius, D., (2005) Scoring Children's Foreign Language Pronunciation, Proceedings, FONETIK 2005, Department of Linguistics, Goteborg University, Sweden
[5] Rudnicky, A. and Xu, W., (1999) An agenda-based dialog management architecture for spoken language systems. IEEE Automatic Speech Recognition and Understanding Workshop, 1999, p I-337.
[6] Боцева, Д., (2008) Честотен речник на лексиката в учебните помагала за деца от 2 до 7 години, Университетско издателство "Климент Охридски", София, България